\documentclass[12pt]{article}

\usepackage{graphics}
\usepackage{epsfig}
\input epsf

\title{On the quark component in prompt photon and electroweak gauge boson 
production at high energies}

\author{A.V.~Lipatov, N.P.~Zotov}

\begin{document}

\maketitle

\begin{center}

{\it D.V.~Skobeltsyn Institute of Nuclear Physics,\\ 
M.V. Lomonosov Moscow State University,
\\119991 Moscow, Russia\/}\\[3mm]

\end{center}

\vspace{0.5cm}

\begin{center}

{\bf Abstract }

\end{center}

In the framework of the $k_T$-factorization approach, 
we study the production of prompt photons and electroweak gauge bosons
in high energy proton-(anti)proton collisions at modern colliders.
Our consideration is based on the amplitude for the production of a 
single photon or $W^{\pm}/Z^0$ boson associated with a quark pair 
in the fusion of two off-shell gluons. The quark component
is taken into account separately using the quark-gluon scaterring and quark-antiquark 
annihilation QCD subprocesses. Special attention is put on the 
contributions from the quarks involved into the earlier steps of the 
evolution cascade. Using the Kimber-Martin-Ryskin formalism, we 
simulate this component and demonstrate that it plays an important 
role at both the Tevatron and LHC energies.
Our theoretical results are compared with recent
experimental data taken by the D$\oslash$ and CDF collaborations 
at the Tevatron.

\vspace{1cm}

\noindent
PACS number(s): 12.38.-t, 12.38.Bx

\vspace{0.5cm}

\section{Introduction} \indent 

The theoretical and experimental studying the prompt 
photon and electroweak gauge boson production at high energies provide
an important information about the nature of both the underlying electroweak
interaction and the effects of Quantum Chromodynamics (QCD).
In many respects these processes have become
one of most important "standard candles" in experimental
high energy physics~[1--11]. 

In the previous publications~[12--14], we have considered the
production of prompt photons and electroweak gauge bosons
$W^\pm$ and $Z^0$ in the $k_t$-factorization approach.
Making use of the $k_T$-factorization
is motivated by the fact that it provides solid theoretical
grounds for the effects of initial gluon radiation and intrinsic
parton transverse momentum $k_T$. We pay
attention to individual contributions from the different
partonic subprocesses. The idea of~[12, 14]  was in reexpressing the quark
contributions in terms of gluon contributions, thus reducing the
problem of poorly known and poorly calculable unintegrated quark
densities to much better investigated gluon ones.
Our studies, however, reveal the fact that the non-reducible quark
distributions are of major importance for the
processes under study. The goal of this paper is to 
clarify this point in more detail and to
accomplish the calculations presented in~[12, 14] by including the contributions
which yet have not been taken into account.

The outline of our paper is following. In Section~2 we 
recall shortly the basic formulas of the $k_T$-factorization approach with a brief 
review of calculation steps. In Section~3 we present the numerical results
of our calculations. The central point is discussing the role
of each contribution to the cross sections.
Section~4 contains our conclusions.

\section{Theoretical framework}
\subsection{The subprocesses under consideration} \indent 

Our approach is the following.
The starting point of consideration is the leading order ${\cal O}(\alpha)$ and ${\cal O}(\alpha \alpha_s)$ subprocesses:
$q + g^* \to \gamma + q$, $q + \bar q \to \gamma + g$ and $q + \bar{q}' \to W^\pm/Z^0$. 
These subprocesses are strongly depend on the unintegrated quark distributions in a proton $f_q(x,{\mathbf k}_T^2,\mu^2)$.
In contrast to the study~[15] where the hard matrix elements of these subprocesses have been
convoluted with the relevant unintegrated quark and/or gluon distributions in a proton,
we try to reexpress the unintegrated quark densities in terms of gluon ones.
Our main idea is connected with the separation of the unintegrated 
quark distributions into several parts which correspond to the interactions of 
valence quarks $f_q^{(v)}(x,{\mathbf k}_T^2,\mu^2)$,
sea quarks appearing at the last step of the gluon evolution $f_q^{(g)}(x,{\mathbf k}_T^2,\mu^2)$ 
and sea quarks coming from the earlier gluon 
splittings $f_q^{(s)}(x,{\mathbf k}_T^2,\mu^2)$ (see Fig.~1).
In our approach, we simulate the 
last gluon splittings by the higher-order ${\cal O}(\alpha \alpha_s^2)$ off-shell (i.e. $k_T$-dependent) matrix elements,
namely $g^* + g^* \to \gamma/W^\pm/Z^0 + q + \bar{q}'$.
In this way we take into account the contributions from the $f_q^{(g)}(x,{\mathbf k}_T^2,\mu^2)$.
To estimate the contributions from the $f_q^{(v)}(x,{\mathbf k}_T^2,\mu^2)$ and $f_q^{(s)}(x,{\mathbf k}_T^2,\mu^2)$ 
we use the specific properties of the Kimber-Martin-Ryskin (KMR) scheme~[16]
which enables us to discriminate between the various components of the
unintegrated quark densities\footnote{Below we will refer to the $f_q^{(s)}(x,{\mathbf k}_T^2,\mu^2)$ contribution as to "reduced sea" component.}. 
Thus, the proposed scheme results to the following partonic subprocesses:
$$
  g^* + g^* \to \gamma/W^\pm/Z^0 + q + \bar{q}', \eqno(1)
$$
$$
  q^{(v)} + g^* \to \gamma/W^\pm/Z^0 + q', \eqno(2)
$$
$$
  q^{(s)} + g^* \to \gamma/W^\pm/Z^0 + q', \eqno(3)
$$
$$
  q + \bar q \to \gamma + g, \quad q + \bar{q}'\to W^\pm/Z^0.\eqno(4)
$$

\noindent
To be precise, the gluon-gluon fusion subprocess~(1) replaces the 
$q^{(g)} + \bar{q}^{(g)}$ annihilation, and the valence and sea quark-gluon 
scattering~(2) and~(3) replace the $q^{(v)} + \bar{q}^{(g)}$ and $q^{(s)} + \bar{q}^{(g)}$ annihilation mechanisms.
In the last two cases both valence $q^{(v)}$ and "reduced sea" $q^{(s)}$ quark components are included.
Of course, all incoming gluons in~(1) --- (4) are off-shell.
To avoid the double counting we have not considered here
$q + \bar q^\prime \to W^\pm/Z^0 + g$ subprocess.

As it was mentioned above, the proposed scheme was applied already to the
prompt photon~[12] and electroweak boson~[14] production at the Tevatron and LHC energies.
However, in~[12] we have neglected the "reduced sea" contribution.
In~[14], to estimate the "reduced sea" component only the contribution from 
the $q^{(s)} + \bar q^{(v)} \to W^\pm/Z^0$ subprocess has been taken into account
since this contribution is the dominant one at the Tevatron energies (see also discussion in Section~3). 
In the present study we give a more accurate analysis of all possible 
contributions to the cross sections of processes under consideration\footnote{In the 
calculations of the prompt photon cross sections we will neglect the 
contributions from the so-called fragmentation mechanism~[17]. It is because after applying the 
isolation cut (see~[7--11]) these contributions amount only to about 10\% of the visible cross 
section. The isolation requirement and additional conditions which 
preserve our calculations from divergences have been specially discussed in~[12, 13].}
and clarify the role of missing contributions to the "reduced sea" component at the LHC.

\subsection{Cross section for the inclusive $\gamma/W^\pm$/$Z^0$ production} \indent 

To calculate the cross section of the prompt photon and/or electroweak boson production
in the framework of the $k_T$-factorization approach one should convolute the 
off-shell matrix elements of subprocesses (1) --- (4) with the
relevant unintegrated quark and/or gluon distributions.
The contribution to the inclusive $\gamma/W^\pm$/$Z^0$ production cross section 
from the off-shell gluon-gluon fusion (1) can be written as
$$
  \displaystyle \sigma(p + \bar p \to \gamma/W^\pm/Z^0 + X) = \sum_{q} \int {|\bar {\cal M}(g^* + g^* \to \gamma/W^\pm/Z^0 + q + \bar q^\prime)|^2\over 256\pi^3 (x_1 x_2 s)^2} \times \atop 
  \displaystyle \times f_g(x_1,{\mathbf k}_{1 T}^2,\mu^2) f_g(x_2,{\mathbf k}_{2 T}^2,\mu^2) d{\mathbf k}_{1 T}^2 d{\mathbf k}_{2 T}^2 d{\mathbf p}_{1 T}^2 {\mathbf p}_{2 T}^2 dy dy_1 dy_2 {d\phi_1\over 2\pi} {d\phi_2\over 2\pi} {d\psi_1\over 2\pi} {d\psi_2\over 2\pi}, \eqno(5)
$$

\noindent
where $|\bar {\cal M}(g^* + g^* \to \gamma/W^\pm/Z^0 + q +\bar{q}')|^2$ is 
the off-mass shell matrix element squared (and averaged over the initial 
gluon polarizations and colors); $\sqrt s$ is the total energy of the process 
under consideration; ${\mathbf k}_{1 T}$, ${\mathbf k}_{2 T}$, $\phi_1$ and $\phi_2$ are 
the transverse momenta and azimuthal angles of the initial off-shell gluons (having the fractions
$x_1$ and $x_2$ of the incoming protons longitudinal momenta);
${\mathbf p}_T$ and $y$ are the transverse momentum and rapidity of the produced prompt photon or vector boson;
${\mathbf p}_{1T}$ and ${\mathbf p}_{2T}$ the transverse momenta of the co-produced quark and antiquark; 
$y_1$, $y_2$, $\psi_1$ and $\psi_2$ are the quark rapidities and azimuthal angles, respectively.
The formulas for the partonic subprocesses (2) --- (4) are 
similar and can be written as follows:
$$
  \displaystyle \sigma(p + \bar p \to \gamma/W^\pm/Z^0 + X) = \sum_{q} \int {|\bar {\cal M}(q + g^* \to \gamma/W^\pm/Z^0 + q^\prime)|^2\over 16\pi (x_1 x_2 s)^2} \times \atop 
  \displaystyle \times f_q(x_1,{\mathbf k}_{1 T}^2,\mu^2) f_g(x_2,{\mathbf k}_{2 T}^2,\mu^2) d{\mathbf k}_{1 T}^2 d{\mathbf k}_{2 T}^2 d{\mathbf p}_{T}^2 dy dy^\prime {d\phi_1\over 2\pi} {d\phi_2\over 2\pi}, \eqno(6)
$$
$$
  \displaystyle \sigma(p + \bar p \to \gamma + X) = \sum_{q} \int {|\bar {\cal M}(q + \bar q\to \gamma + g|^2\over 16\pi (x_1 x_2 s)^2} \times \atop 
  \displaystyle \times f_q(x_1,{\mathbf k}_{1 T}^2,\mu^2) f_q(x_2,{\mathbf k}_{2 T}^2,\mu^2) d{\mathbf k}_{1 T}^2 d{\mathbf k}_{2 T}^2 d{\mathbf p}_{T}^2 dy dy_g {d\phi_1\over 2\pi} {d\phi_2\over 2\pi}, \eqno(7)
$$
$$
  \displaystyle \sigma(p + \bar p \to W^\pm/Z^0 + X) = \sum_{q} \int {2 \pi \over (x_1 x_2 s)^2} |\bar {\cal M}(q + \bar q^\prime \to W^\pm/Z^0)|^2 \times \atop 
  \displaystyle \times f_q(x_1,{\mathbf k}_{1 T}^2,\mu^2) f_q(x_2,{\mathbf k}_{2 T}^2,\mu^2) d{\mathbf k}_{1 T}^2 d{\mathbf k}_{2 T}^2 dy {d\phi_1\over 2\pi} {d\phi_2\over 2\pi}, \eqno(8)
$$

\noindent
where $y^\prime$ and $y_g$ are the rapidities of the final quark or gluon.
The analytic expressions for the off-shell matrix elements of subprocesses (1) --- (4)
has been derived in our previous papers~[12, 14] (see also~[18]).
We only mention here that, in accord with the $k_T$-factorization
prescription~[19, 20], the off-shell gluon spin density matrix has been 
taken in the form
$$
  \sum \epsilon^\mu (k_i) \epsilon^{*\,\nu} (k_i) = { k_{iT}^\mu k_{iT}^\nu \over {\mathbf k}_{iT}^2}. \eqno(9)
$$

\noindent 
In all other respects our calculations follow the standard Feynman rules.
If we average these expressions over $\phi_{1}$ and $\phi_{2}$ 
and take the limit ${\mathbf k}_{1 T}^2 \to 0$ and ${\mathbf k}_{2 T}^2 \to 0$,
then we recover the relevant formulas in the leading order collinear approximation of QCD.

\subsection{The KMR unintegrated parton distributions} \indent 

In further analysis below we will use the unintegrated quark and gluon densities in a proton
which taken in the KMR form~[16].
The KMR approach is the formalism to construct the unintegrated parton distributions
$f_a(x,{\mathbf k}_T^2,\mu^2)$ from the known conventional parton
distributions $xa(x,\mu^2)$, where $a = g$ or $a = q$. 
In this approximation, the unintegrated quark and 
gluon distributions are given by~[16]
$$
  \displaystyle f_q(x,{\mathbf k}_T^2,\mu^2) = T_q({\mathbf k}_T^2,\mu^2) {\alpha_s({\mathbf k}_T^2)\over 2\pi} \times \atop {
  \displaystyle \times \int\limits_x^1 dz \left[P_{qq}(z) {x\over z} q\left({x\over z},{\mathbf k}_T^2\right) \Theta\left(\Delta - z\right) + P_{qg}(z) {x\over z} g\left({x\over z},{\mathbf k}_T^2\right) \right],} \eqno (10)
$$
$$
  \displaystyle f_g(x,{\mathbf k}_T^2,\mu^2) = T_g({\mathbf k}_T^2,\mu^2) {\alpha_s({\mathbf k}_T^2)\over 2\pi} \times \atop {
  \displaystyle \times \int\limits_x^1 dz \left[\sum_q P_{gq}(z) {x\over z} q\left({x\over z},{\mathbf k}_T^2\right) + P_{gg}(z) {x\over z} g\left({x\over z},{\mathbf k}_T^2\right)\Theta\left(\Delta - z\right) \right],} \eqno (11)
$$

\noindent
where $P_{ab}(z)$ are the usual unregulated LO DGLAP splitting 
functions. The theta functions which appear 
in~(10) and~(11) imply the angular-ordering constraint $\Delta = \mu/(\mu + |{\mathbf k}_T|)$ 
specifically to the last evolution step to regulate the soft gluon
singularities. For other evolution steps, the strong ordering in 
transverse momentum within the DGLAP equations automatically 
ensures angular ordering\footnote{Numerically, in (10) and (11) we have applied the recent Martin-Stirling-Thorne-Watt (MSTW) LO
parametrizations~[21] of the collinear parton densities $a(x,\mu^2)$.
This choice is differs from the one~[12--14] where the Gl\"uck-Reya-Vogt (GRV) parton
distributions~[22] have been used.}.
The Sudakov form factors $T_q({\mathbf k}_T^2,\mu^2)$ and 
$T_g({\mathbf k}_T^2,\mu^2)$ which appears in (10) and (11) enable us to include logarithmic loop corrections
to the calculated cross sections. The nonlogarithmic corrections 
can be taken into account by using the $K$-factor~[15, 23] 
$K(q+\bar{q}'\to W/Z)\simeq \exp\left[C_F \pi\alpha_s(\mu^2)/2\right]$
with $C_F=4/3$ and $\mu^2={\mathbf p}_T^{4/3}m^{2/3}$.

The function $f_q(x,{\mathbf k}_T^2,\mu^2)$ in~(10)
represents the total quark distribution function in a proton.
Modifying~(10) in such a way that only the first term is 
kept and the second term omitted, we switch the last gluon splitting 
off, thus excluding the $f_q^{(g)}(x,{\mathbf k}_T^2,\mu^2)$ component.
Taking the difference between the quark and antiquark densities we
extract the valence quark component $f_q^{(v)}(x,{\mathbf k}_T^2,\mu^2)=
f_q(x,{\mathbf k}_T^2,\mu^2)-f_{\bar{q}}(x,{\mathbf k}_T^2,\mu^2)$.
Finally, keeping only sea quark in first term of~(10) we remove the 
valence quarks from the evolution ladder.
In this way only the $f_q^{(s)}(x,{\mathbf k}_T^2,\mu^2)$ contributions to the 
$f_q(x,{\mathbf k}_T^2,\mu^2)$ 
are taken into account. 

The multidimensional integration in~(5) --- (8) has been performed
by the means of Monte Carlo technique, using the routine 
\textsc{Vegas}~[24]. The full C$++$ code is available from the 
authors on request\footnote{lipatov@theory.sinp.msu.ru}.
This code is practically identical to that used in~[12--14].

\section{Numerical results} \indent

We are now in a position to present our numerical results.
In all our calculations according to (5) --- (8) the light quark masses were set to 
$m_u = 4.5$~MeV, $m_d = 8.5$~MeV and $m_s = 155$~MeV. 
We have checked that the uncertainties coming from these quantities 
are negligible compared to the uncertainties connected with the scales in the 
unintegrated parton densities and strong coupling.
We set $m_c = 1.4$~GeV, $m_W=80.403$~GeV, $m_Z=91.1876$~GeV, $\sin^2 \theta_W=0.23122$
and use the LO formula for the strong 
coupling constant $\alpha_s(\mu^2)$ with $n_f = 4$ 
active quark flavors at $\Lambda_{\rm QCD} = 200$~MeV (so that $\alpha_s(M_Z^2) = 0.1232$).
As it is often done, we choose the renormalization and factorization 
scales to be equal: $\mu_R = \mu_F = \mu = m_T$, where $m_T$ is the transverse mass 
of the produced vector boson. In the case of prompt photon production
we set the scale $\mu$ to be equal to the photon transverse energy $E_T^\gamma$.
We will not study here the scale dependense of our results. 
This issue is addressed in our previous papers~[12--14].

\subsection{Role of the quark contributions} \indent 

We begin the discussion by presenting a comparison
between the different contributions to the $\gamma/W^\pm/Z^0$
cross sections. In Fig.~2 we plot our results for the cross sections as a function of 
produced photon or gauge boson center-of-mass rapidity $y$. Here, we have performed the calculations for both the 
proton-antiproton and proton-proton interactions at the Tevatron and LHC energies, respectively.
In the case of electroweak boson production, the cross sections are
multiplied by the branching fractions $f(W \to l\nu)$ and $f(Z \to l^+l^-)$. We set these
branching fractions to $f(W \to l\nu) = 0.1075$ and $f(Z \to l^+l^-) = 0.03366$~[25].
The additional cuts $|y| < 2.5$ and $|y| < 4$ have been 
applied in the case of prompt photon production at the Tevatron and LHC.
The solid, dashed and dotted histograms in Fig.~2
represent the contributions from the $g^* + g^* \to \gamma/W^\pm/Z^0 + q +\bar{q}'$,
$q_v + g^* \to \gamma/W^\pm/Z^0 + q'$ and $q_v + \bar{q}'_v \to W^\pm/Z^0$ 
(or $q_v + \bar q_v \to \gamma + g$) subprocesses, 
respectively\footnote{In the present analysis the contributions 
from all $2 \to 1$ subprocesses have been corrected by the factor of $3/2$
compared to those from~[14]. We thank V.A.~Saleev 
for drawing our attention to this point.}. 
The thick solid histograms represent the sum of all contributions.
One can see that the gluon-gluon fusion is an important photon production mechanism
at both the Tevatron and LHC conditions.
Moreover, it gives a main contribution to the cross section at the LHC.
In the case of $W^\pm/Z^0$ production, the role of gluon-gluon fusion subprocess
is negligible at the Tevatron and is increased greatly at the LHC energy:
it contributes only about one or two percent to the total cross section at $\sqrt = 1800$~GeV 
and more than 40\% at $\sqrt s = 14$~TeV. 
The Compton-like subprocesses $q_v + g^* \to \gamma/W^\pm/Z^0 + q'$
are also important.

In Fig.~2, the dash-dotted histograms represent the "reduced sea" component.
We find that this component gives approximately 30\% contribution to the
total cross section of prompt photon production at the Tevatron and 
approximately 20\% contribution at the LHC. In the case of electroweak boson production,
it contributes about 50\% and 40\%, respectively.
As it was noted above, this component contains
the $q^{(s)} + g^* \to \gamma/W^\pm/Z^0 + q'$, 
$q^{(s)} + \bar q^{(s)} \to \gamma + g$ (or $q^{(s)} + \bar{q}^{(s)}\to W^\pm/Z^0$)
and $q^{(s)} + \bar q^{(v)} \to \gamma + g$ (or $q^{(s)} + \bar{q}^{(v)}\to W^\pm/Z^0$)
subprocesses. The relative contributions are shown in Fig.~3.
Note that thick solid histograms in Fig.~3 corresponds to the dash-dotted 
histograms in Fig.~2.
Since all these subprocesses 
are mainly due to the quarks emerging from the earlier steps of the parton 
evolution rather than from the last gluon splitting, one can 
conclude that the quarks constitute an important component of the parton ladder,
not negligible even at the LHC energies and not reducible to the gluon
component. 

\subsection{Comparison with the Tevatron data} \indent 

Now we turn to the comparison of our theoretical predictions 
with the experimental data on the prompt photon and $W^\pm/Z^0$ boson
production at Tevatron.
The data~[7--11] on the inclusive prompt photon 
hadroproduction come from both the
D$\oslash$ and CDF collaborations. The D$\oslash$~[7, 8] data 
were obtained in the central and forward pseudo-rapidity
regions for two different center-of-mass energies, namely 
$\sqrt s = 630$ GeV and $\sqrt s = 1800$ GeV.
The central pseudo-rapidity region is defined by the requirement
$|\eta^\gamma| < 0.9$, and the forward one is 
defined by $1.6 < |\eta^\gamma| < 2.5$.
The more recent CDF data~[9, 10] refer to the same central kinematical region
$|\eta^\gamma| < 0.9$  for both beam energies $\sqrt s = 630$ GeV and $\sqrt s = 1800$ GeV.
The data on the electroweak gauge boson production are also come
from the D$\oslash$~[2--6] and CDF~[1] collaborations.

The results of our calculations are shown in Figs.~4 --- 9. 
Fig.~4 confronts the double differential cross sections 
$d\sigma/dE_T d \eta$ of the prompt photon production 
calculated at $\sqrt s = $~630 and 1800~GeV
in different kinematical regions with the D$\oslash$~[7, 8] and CDF~[9, 10] data.
The solid histograms represent calculations 
in the scheme described above.
For comparison, we also show (as dashed histograms) the predictions
based on the simple $2\to 2$ QCD subprocesses with all quark components summed together.
One can see that the both approaches give the very similar results
which agree reasonably with the Tevatron data on the prompt photon cross sections 
within the experimental uncertainties.
This fact demonstrates that the high-order corrections for prompt 
photon production connected with the terms not containing large logarithms 
are rather small. 
Our predictions based on the subprocesses (1) --- (4) are rather 
similar to ones~[26] based on the collinear QCD factorization with the NLO accuracy.

Concerning the electroweak gauge boson production,
the situation is slightly different.
Figs.~5 and~7 display a comparison between the calculated differential cross sections 
$d\sigma/dp_T$ and the D$\oslash$ and the CDF experimental data~[1, 3, 4] at low $p_T$ ($p_T < 20$~GeV) 
and in the full $p_T$ range. These data have been obtained at $\sqrt s = 1800$~GeV.
In Figs.~6, 8 and 9, we show the normalized differential cross section
$(1/\sigma)\,d\sigma/dp_T$ and $(1/\sigma)\,d\sigma/d|y|$ of the $W^\pm$ and $Z^0$ boson
 production in comparison with the data.
The predictions based on the scheme (1) --- (4) 
are about a factor of 1.25 higher than the ones based on the simple $2\to 1$ subprocesses.
The main difference between the predictions is observed in the low $p_T$ region.
This difference can be attributed to the terms not 
containing large logarithms and connected with using of the high-order off-shell 
matrix elements mentioned above.
In contrast with the prompt photon production, 
such terms are significant for the case of $W^\pm/Z^0$ production. 
In~[15], an additional factor of about 1.2 was introduced {\sl ad hoc}
to eliminate the visible disagreement between the data and 
theory. The origin of 
this extra factor has explained~[15] by the fact that the input
parton densities (used to determine the unintegrated ones)
should themselves be determined from data using
the appropriate non-collinear formalism.
The results of our calculations based on the scheme (1) --- (4) show no need in this extra factor. 
Some overestimation of our predictions over experimental data at small $p_T$
can be, in principle, connected with problem of applicability of BFKL-like evolution
in this region and requires an additional study.
Note that the traditional QCD calculations (valid in a wide $p_T$ range) combine fixed-order
perturbation theory (at present, up to NNLO terms~[27--31]) with analytic soft-gluon resummation~[32,33] 
and some matching criterion. These calculations gives a similar description of the data~[1--6].

Additionally we have studied the effects
of the non-logarithmic loop corrections to the gauge boson production amplitude.
To do this, we have repeated the calculations based on the $2\to 1$ quark-antiquark annihilation 
with the omitted $K$-factor. The dotted histograms in Figs.~4 --- 9
correspond to the results of these calculations.
We have found a significant (by a factor of about 1.5) reduction of the predicted
cross sections. Also, in our numerical calculations we have tested the rather old 
GRV (LO) parametrizations~[22] of the collinear parton distributions in a proton (not shown 
in plots). We have found that the difference between the results based on the GRV and recent MSTW~[21]
collinear parton densities is negligible.

Finally, we would like to mention that an additional possibility 
to distinguish the two calculation schemes comes from 
studying the ratio of the $W^\pm$ and $Z^0$ boson 
cross sections. In fact, since $W^\pm$ and $Z^0$ production
properties are very similar, as
the transverse momentum of the vector boson becomes
smaller, the radiative corrections affecting the individual distributions and 
the cross sections of hard process are
factorized and cancelled in this ratio.
Therefore the results of calculation of this ratio 
in the scheme (1) --- (4) (where
the ${\cal O}(\alpha \alpha_s)$ and ${\cal O}(\alpha \alpha_s^2)$ 
subprocesses are taken into account) and the predictions 
based on the ${\cal O}(\alpha)$
quark-antiquark annihilation should
differ from each other at moderate and high $p_T$ values.
This issue have been studied in our previous paper~[14].

\section{Conclusions} \indent 

We have studied the production of prompt photon and
electroweak gauge bosons in hadronic collisions at high energies
in the $k_T$-factorization approach of QCD.
The central part of our consideration is the off-shell gluon-gluon 
fusion subprocess $g^* + g^* \to \gamma/W^\pm/Z^0 + q + \bar q^\prime$. 
The contribution from the quarks has been taken into account additionally.

To study the individual contributions from the different partonic subprocesses
we have used the KMR scheme.
We find that the gluon-gluon fusion is an important production mechanism
of prompt photons at both the Tevatron and LHC conditions.
At the LHC, it gives the main contribution to the cross section.
In the case of $W^\pm/Z^0$ production, it contributes only 
about one or two percent to the total cross section at the 
Tevatron and more than 20\% at the LHC energy.

We demonstrate that an important contribution to the
total cross sections of the processes under consideration also
comes from the sea quark interactions. Notably, we found that the contribution of 
these subprocesses 
are mainly due to the quarks emerging from the earlier steps of the parton 
evolution rather than from the last gluon splitting. 

\section*{Acknowledgements} \indent 

We thank S.P.~Baranov for participation on the initial stage of
investigations, H.~Jung, M.~Deak and F.~Schwennsen for their 
encouraging interest and very helpful discussions.
The authors are very grateful to 
DESY Directorate for the support in the 
framework of Moscow --- DESY project on Monte-Carlo
implementation for HERA --- LHC.
A.V.L. was supported in part by the grants of the president of 
Russian Federation (MK-438.2008.2) and Helmholtz --- Russia
Joint Research Group.
Also this research was supported by the 
FASI of Russian Federation (grant NS-1456.2008.2)
and the RFBR fundation (grant 08-02-00896-a).

\newpage

\begin{figure}
\begin{center}
\epsfig{figure=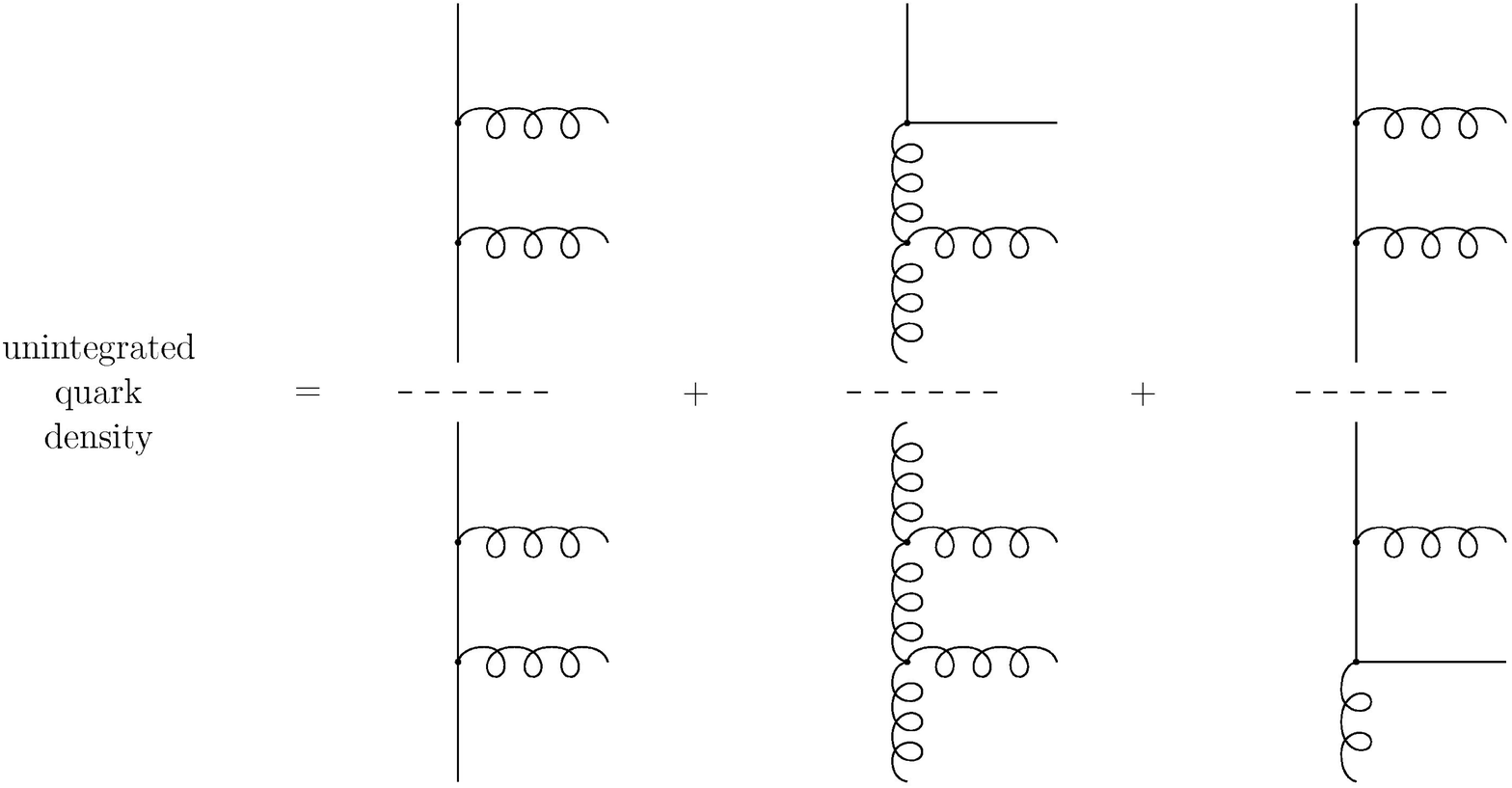, width = 15cm}
\caption{Our approach to calculate the unintegrated quark distributions.}
\end{center}
\label{figure1}
\end{figure}

\newpage

\begin{figure}
\begin{center}
\epsfig{figure=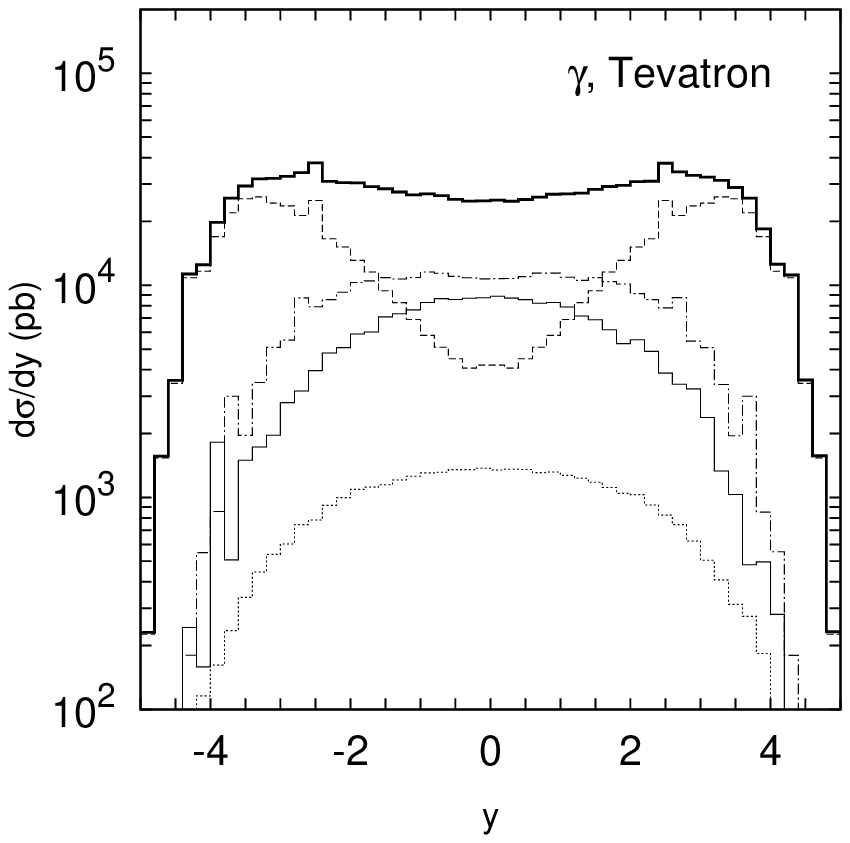, width = 8.1cm}
\epsfig{figure=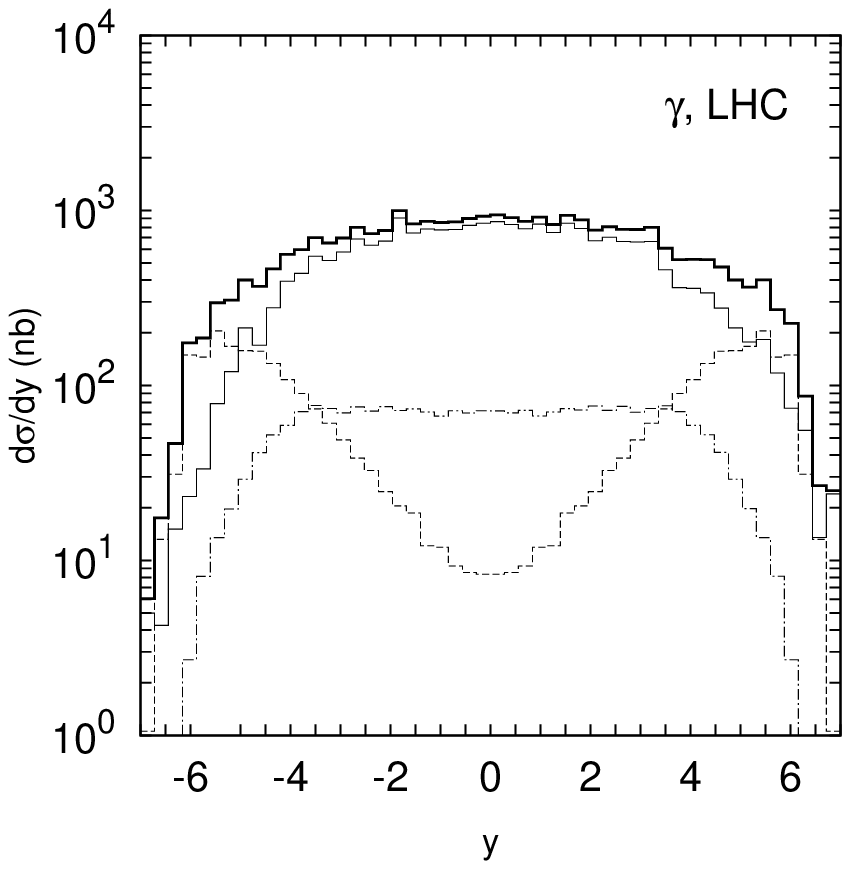, width = 8.1cm}
\epsfig{figure=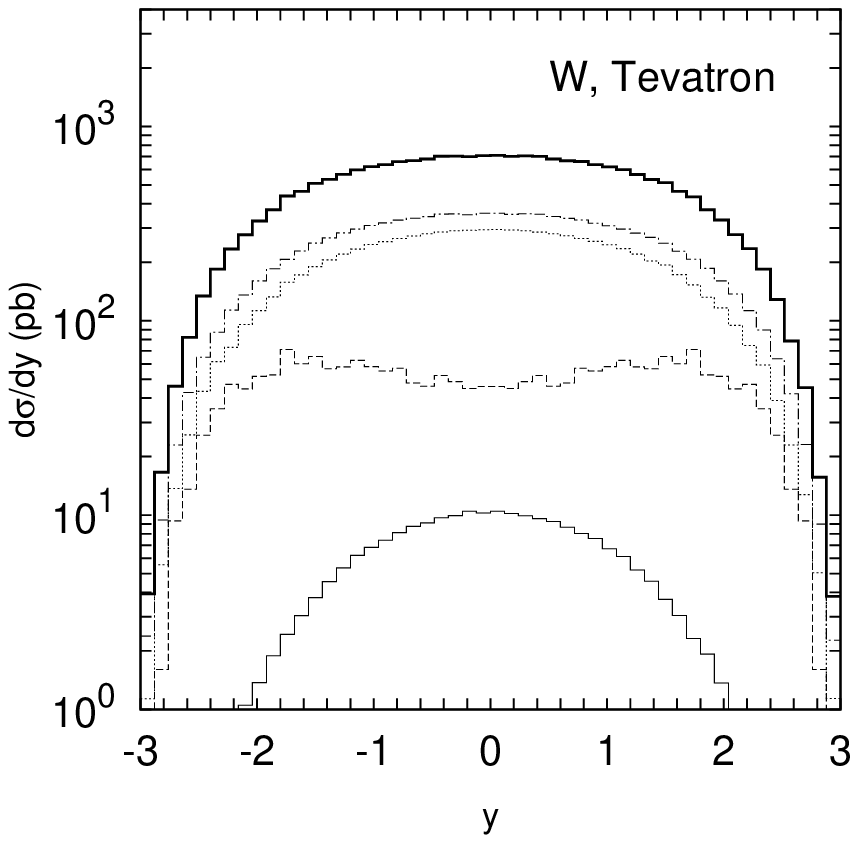, width = 8.1cm}
\epsfig{figure=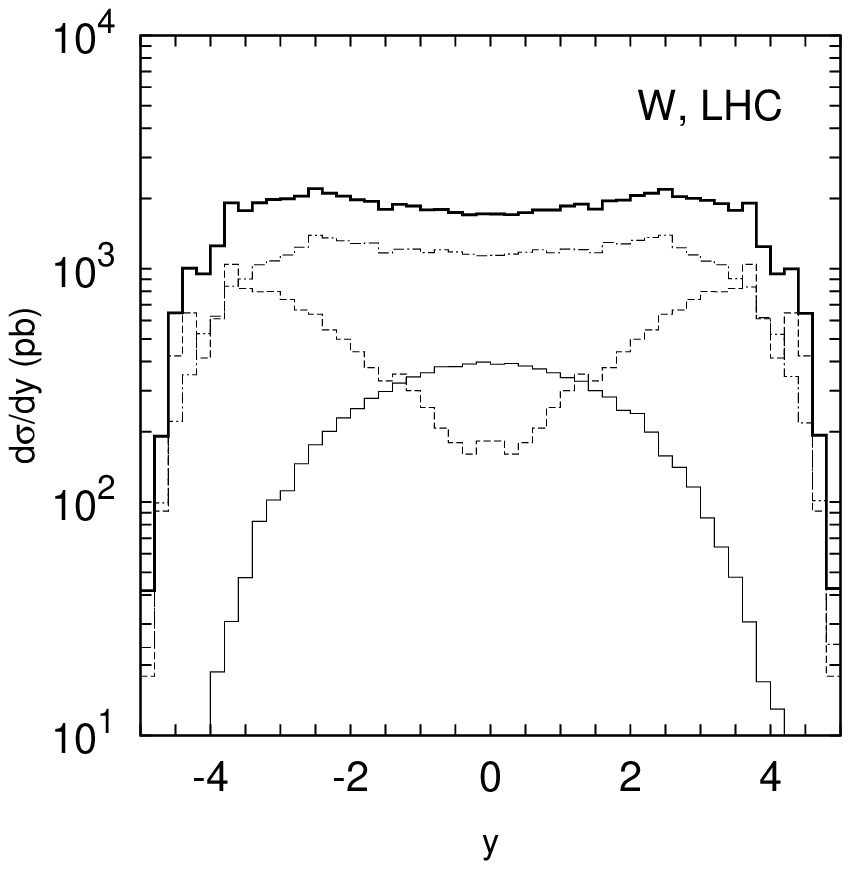, width = 8.1cm}
\epsfig{figure=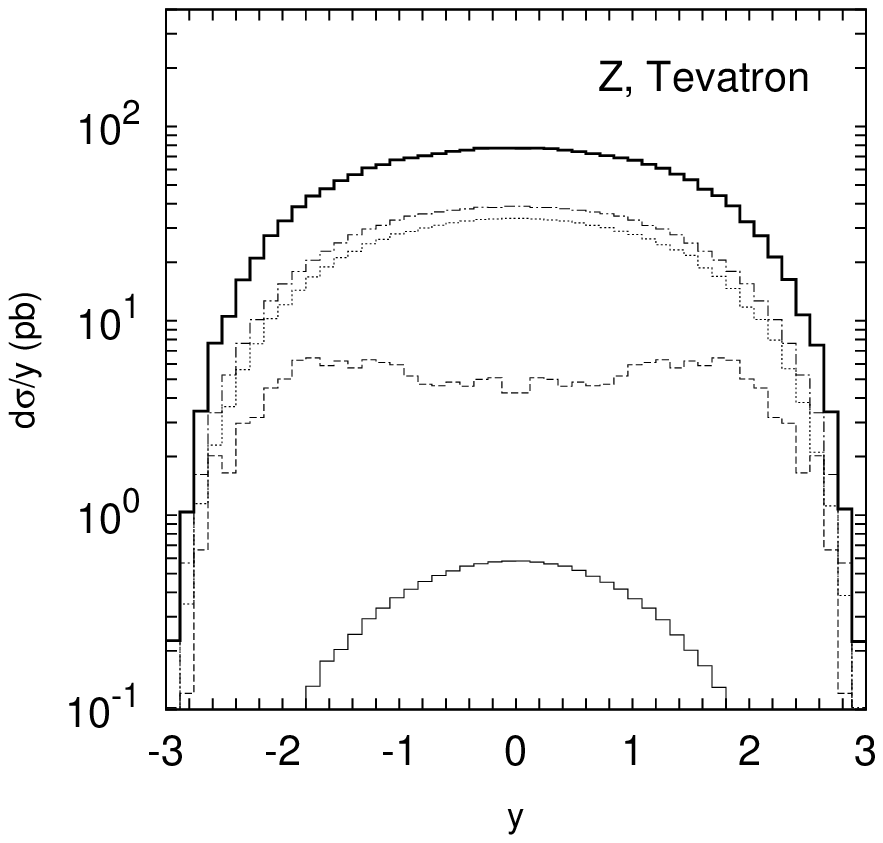, width = 8.1cm}
\epsfig{figure=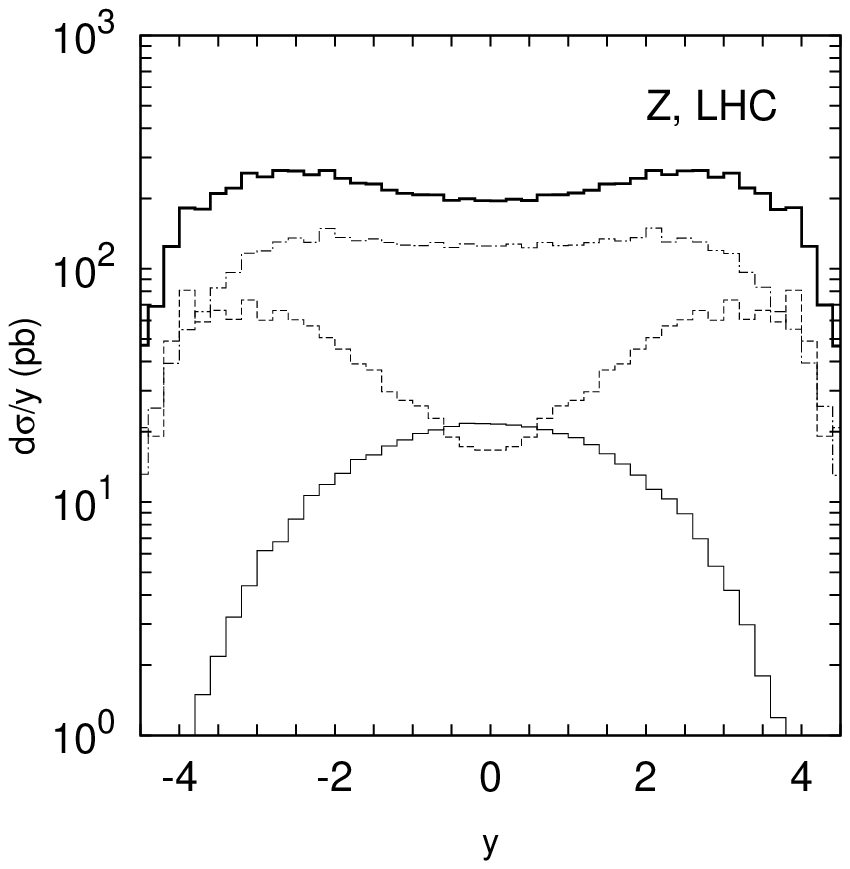, width = 8.1cm}
\caption{Differential cross sections of prompt photon or electroweak boson
production at the Tevatron and LHC as a function of their c.m. rapidity $y$. 
The solid, dashed and dotted histograms
represent the contributions from the $g^* + g^* \to \gamma/W^\pm/Z^0 + q +\bar{q}'$,
$q_v + g^* \to \gamma/W^\pm/Z^0 + q'$ and $q_v + \bar q_v \to \gamma + g$ (or $q_v + \bar q_v \to W^\pm/Z^0$) subprocesses, respectively. 
The dash-dotted histograms represent the "reduced sea" component. 
The thick solid histograms represent the sum of all contributions.}
\end{center}
\label{figure2}
\end{figure}

\newpage

\begin{figure}
\begin{center}
\epsfig{figure=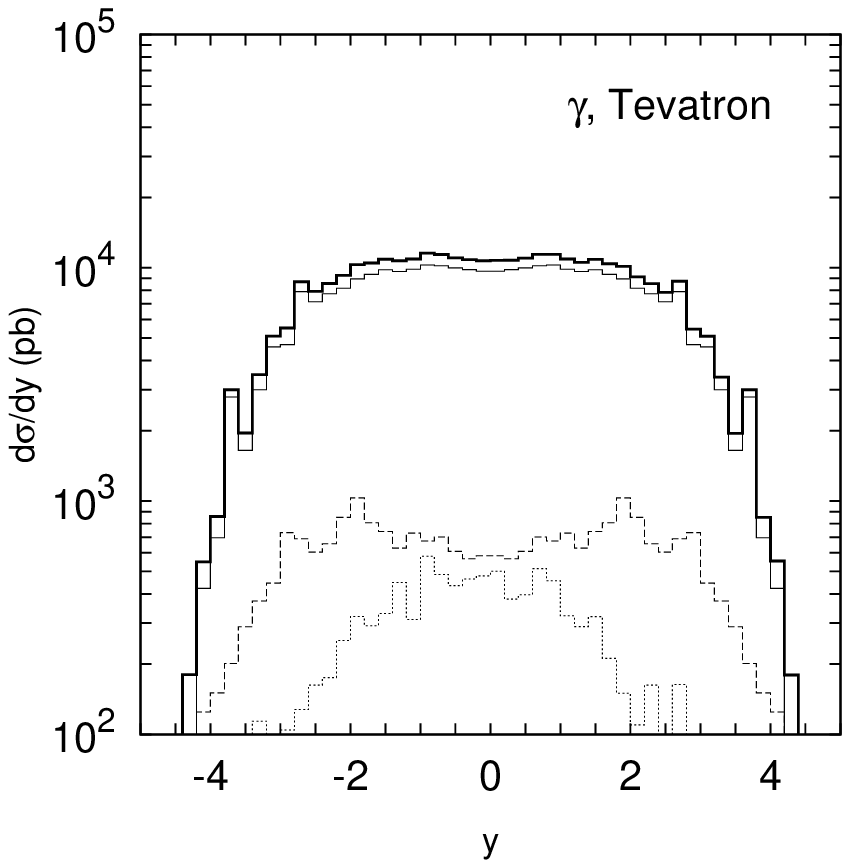, width = 8.1cm}
\epsfig{figure=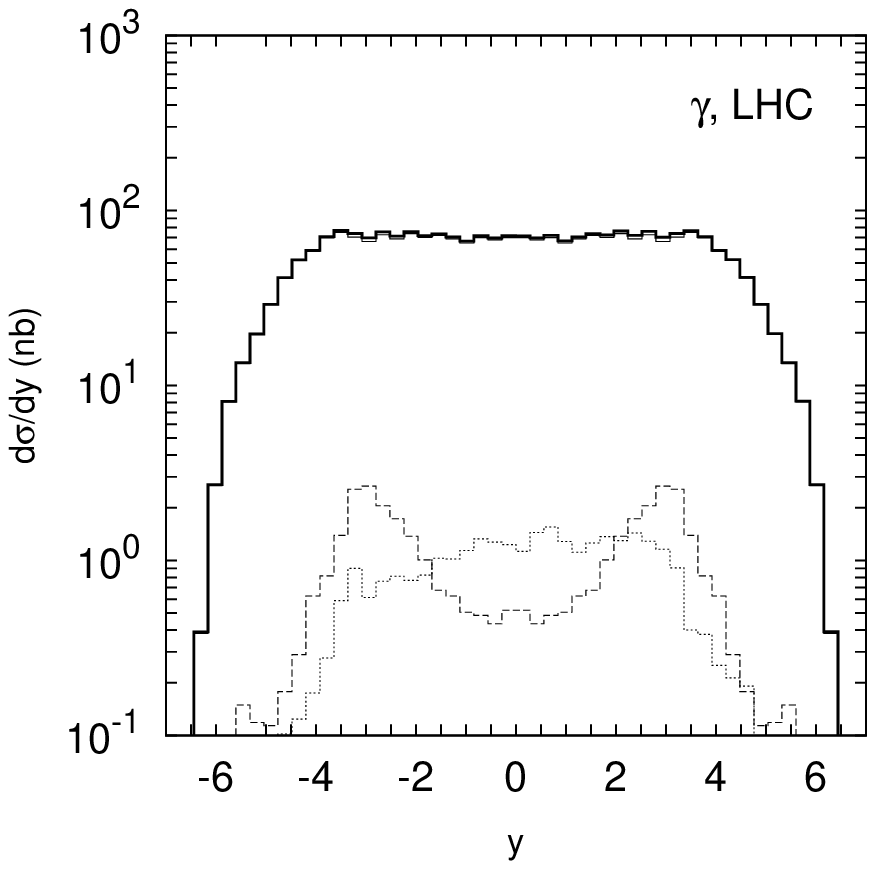, width = 8.1cm}
\epsfig{figure=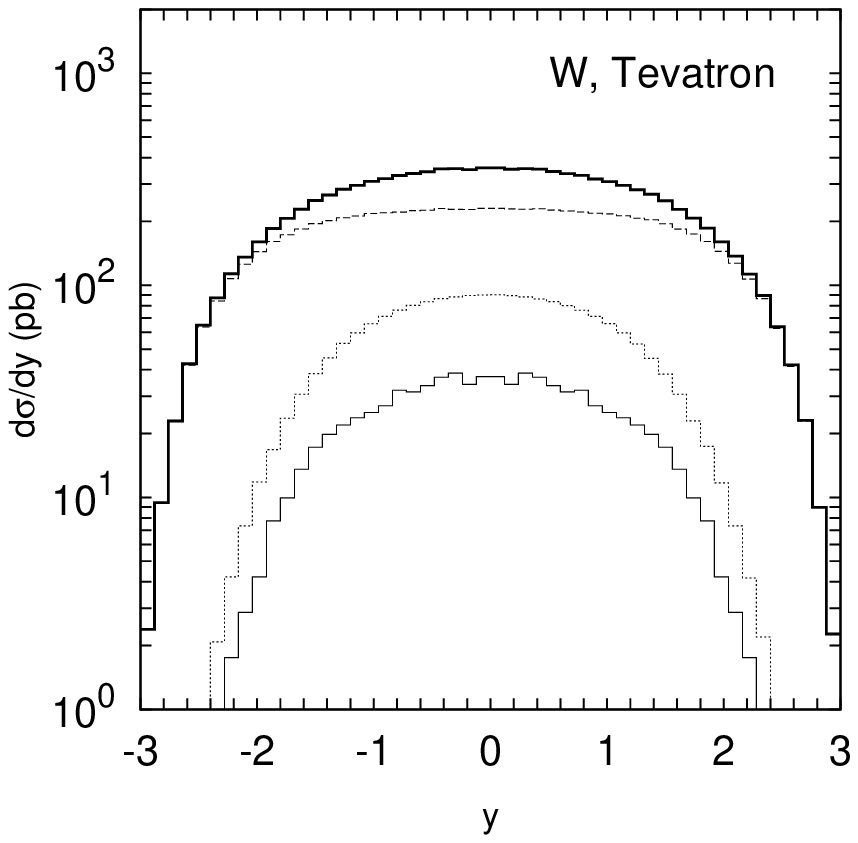, width = 8.1cm}
\epsfig{figure=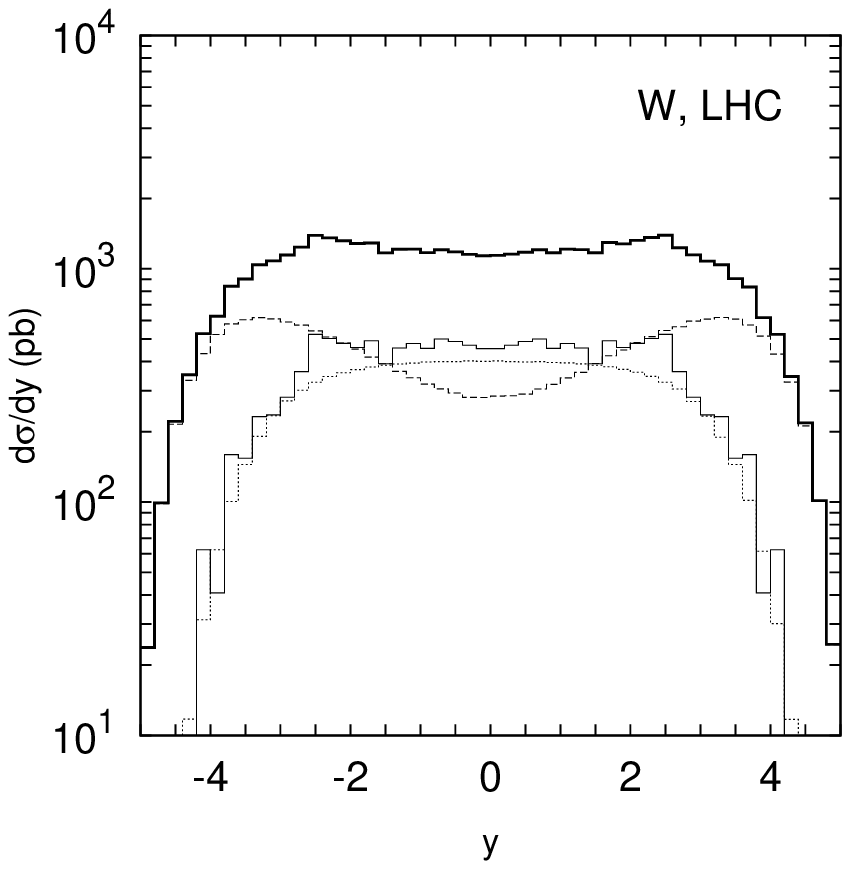, width = 8.1cm}
\epsfig{figure=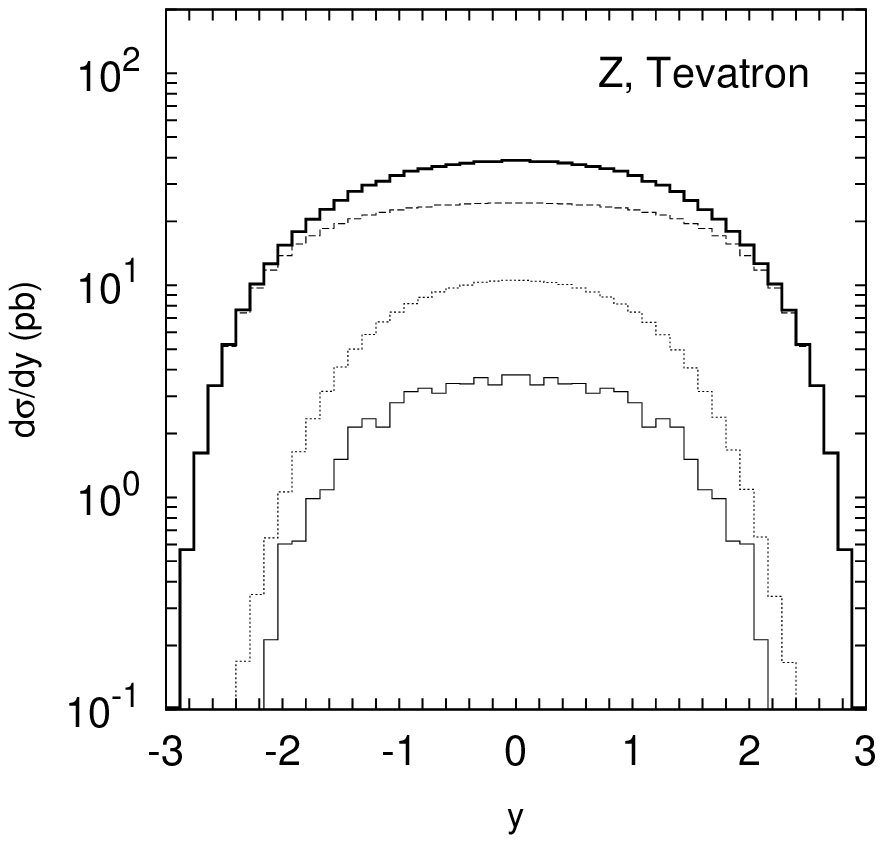, width = 8.1cm}
\epsfig{figure=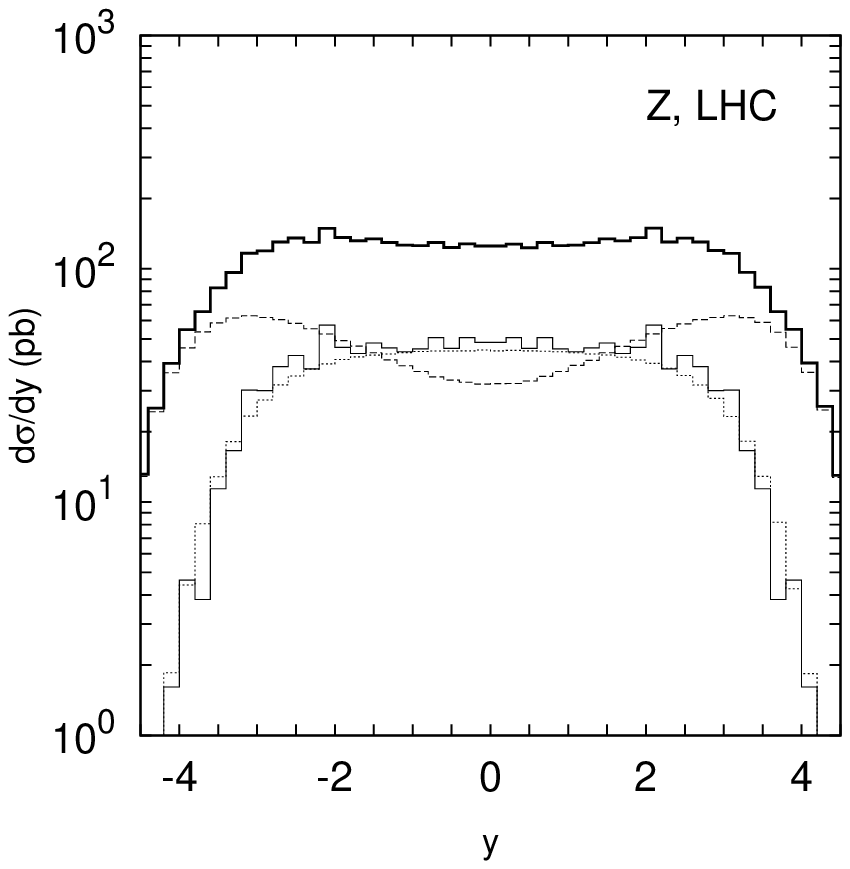, width = 8.1cm}
\caption{Different contributions from the "reduced sea" component to the 
prompt photon or electroweak boson production cross sections at the Tevatron and
LHC. The solid, dashed and dotted histograms
represent the contributions from the $q_s + g^* \to \gamma/W^\pm/Z^0 + q'$, 
$q_s + \bar q_v \to \gamma + g$ (or $q_s + \bar{q}'_v \to W^\pm$) and
$q_s + \bar q_s \to \gamma/W^\pm/Z^0 + g$ subprocesses, respectively. 
The thick solid histograms represent the sum of all contributions.}
\end{center}
\label{figure3}
\end{figure}

\newpage

\begin{figure}
\begin{center}
\epsfig{figure=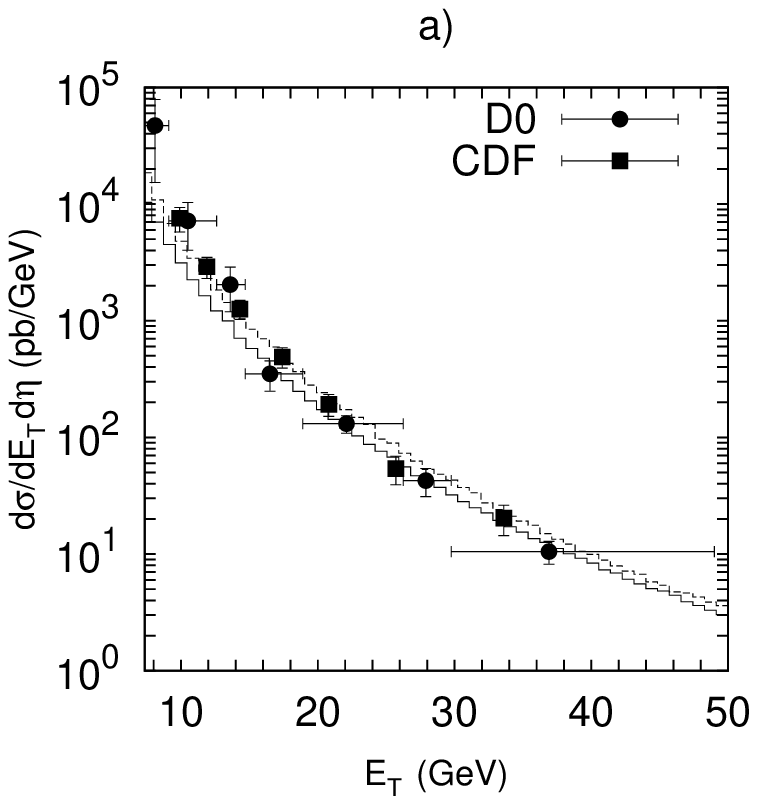, width = 8.1cm}
\epsfig{figure=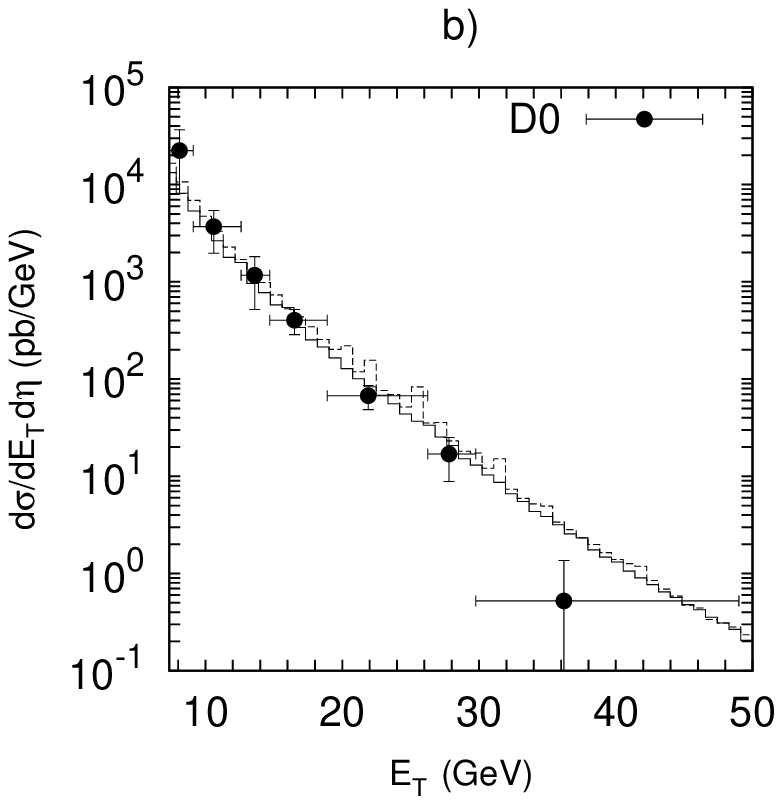, width = 8.1cm}
\epsfig{figure=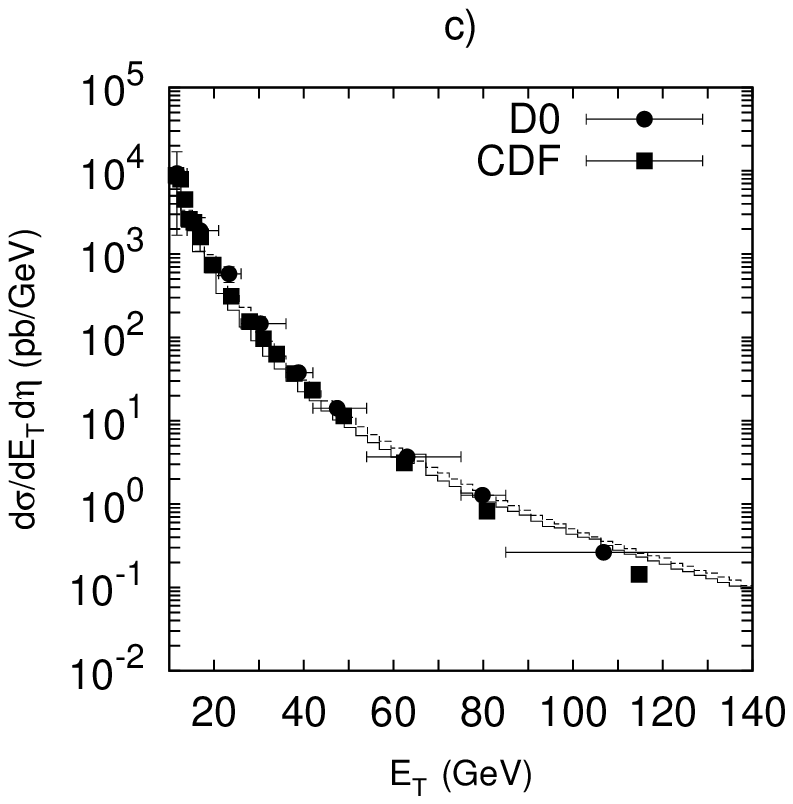, width = 8.1cm}
\epsfig{figure=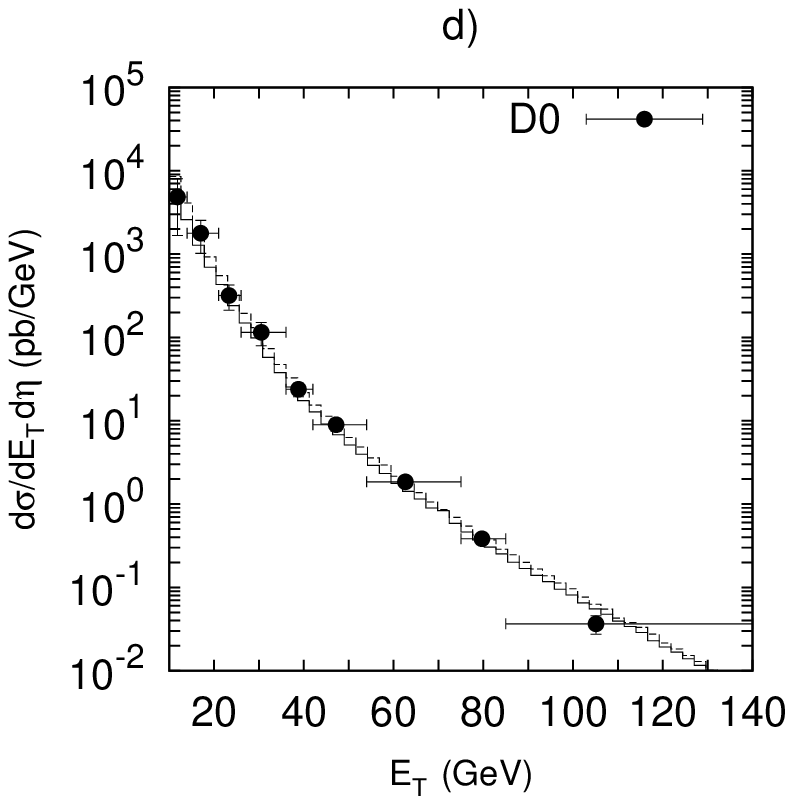, width = 8.1cm}
\caption{Double differential cross section $d\sigma/d E_T d\eta$
for the inclusive prompt photon hadroproduction calculated for 
$|\eta| < 0.9$ (a) and $1.6 < |\eta| < 2.5$ (b) at $\sqrt s = 630$ GeV and 
for $|\eta| < 0.9$ (c) and $1.6 < |\eta| < 2.5$ (d) at $\sqrt s = 1800$ GeV.
Solid histograms represent calculations in the "decomposition" scheme
where all contributions described in the text are taken into account.
Dashed histograms correspond to the predictions based on the simple $2\to 2$
quark-gluon QCD interaction and quark-antiquark annihilation subprocess with all 
quark components summed together.
The experimental data are from D$\oslash$~[7, 8] and CDF~[9, 10].}
\end{center}
\label{figure4}
\end{figure}

\newpage

\begin{figure}
\begin{center}
\epsfig{figure=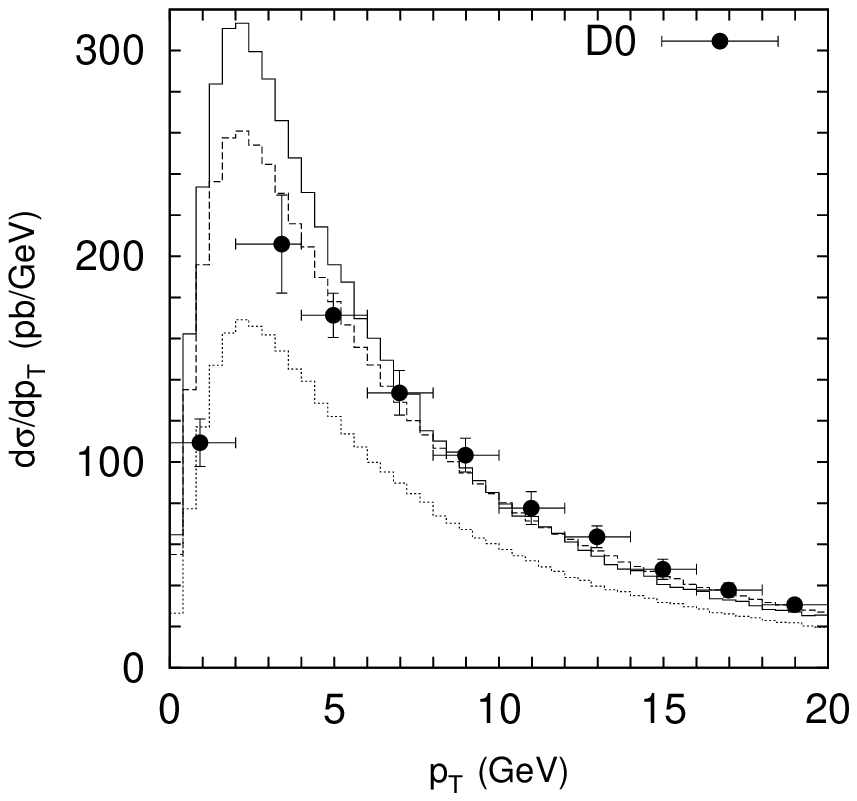, width = 8.1cm}
\epsfig{figure=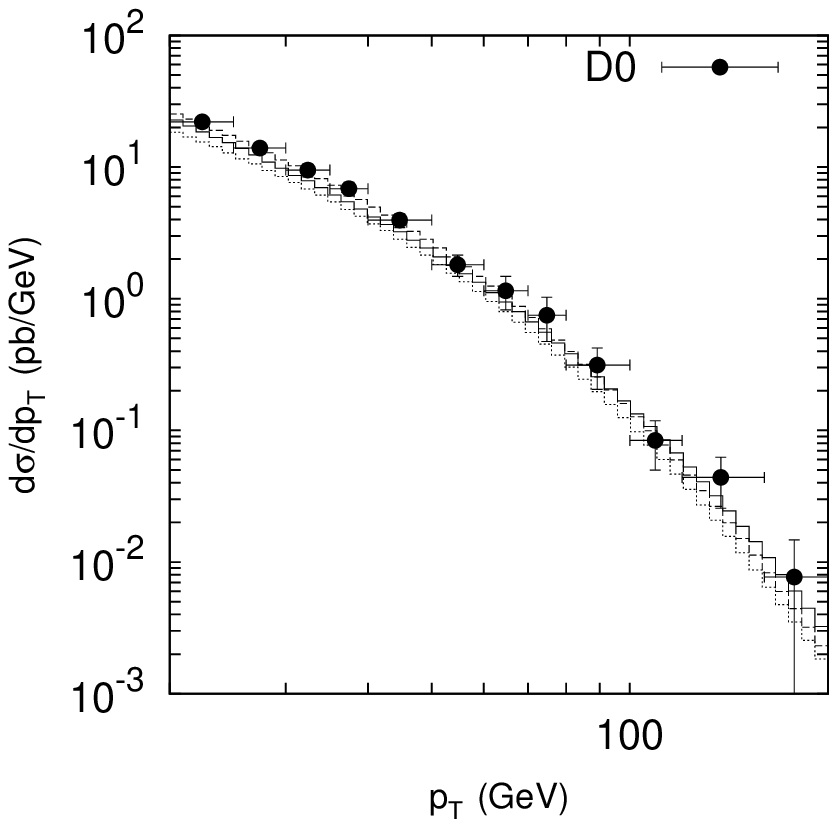, width = 8.1cm}
\caption{Transverse momentum distribution of
the $W^\pm$ boson production calculated at $\sqrt s = 1800$~GeV. 
Solid histograms represent calculations in the "decomposition" scheme
where all contributions described in the text are taken into account.
Dashed histograms correspond to the predictions based on the simple $2\to 1$
quark-antiquark annihilation subprocess with all quark components summed together.
Dotted histograms correspond to the simple $2\to 1$
quark-antiquark annihilation subprocess without $K$-factor.
The cross sections time branching fraction $f(W \to l\nu)$ are shown.
The experimental data are from D$\oslash$~[4].}
\end{center}
\label{figure5}
\end{figure}

\newpage

\begin{figure}
\begin{center}
\epsfig{figure=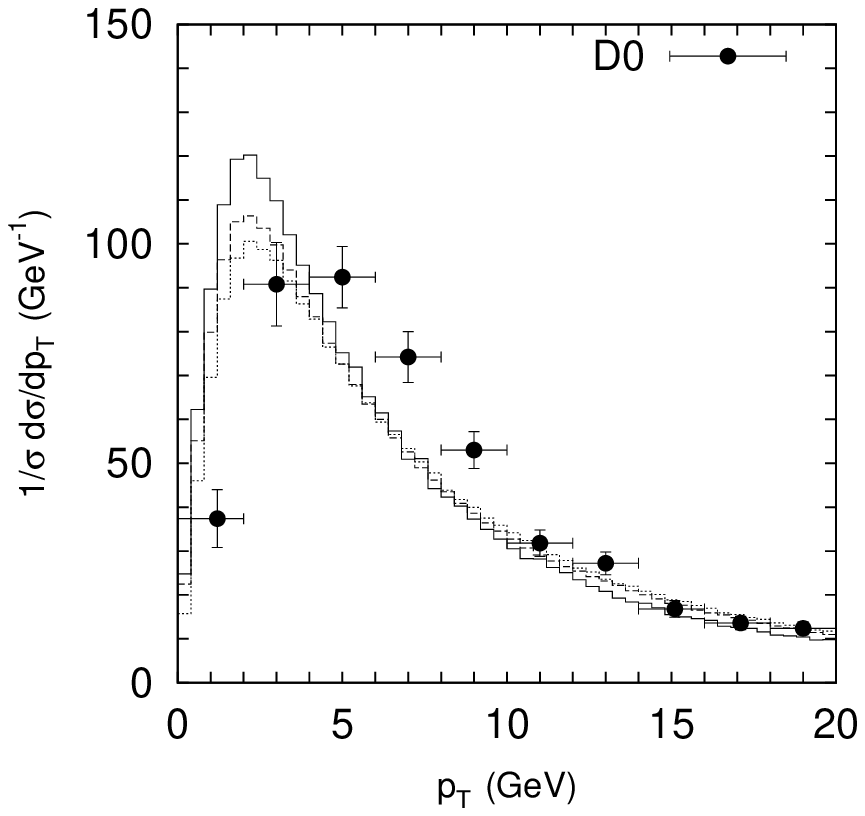, width = 8.1cm}
\epsfig{figure=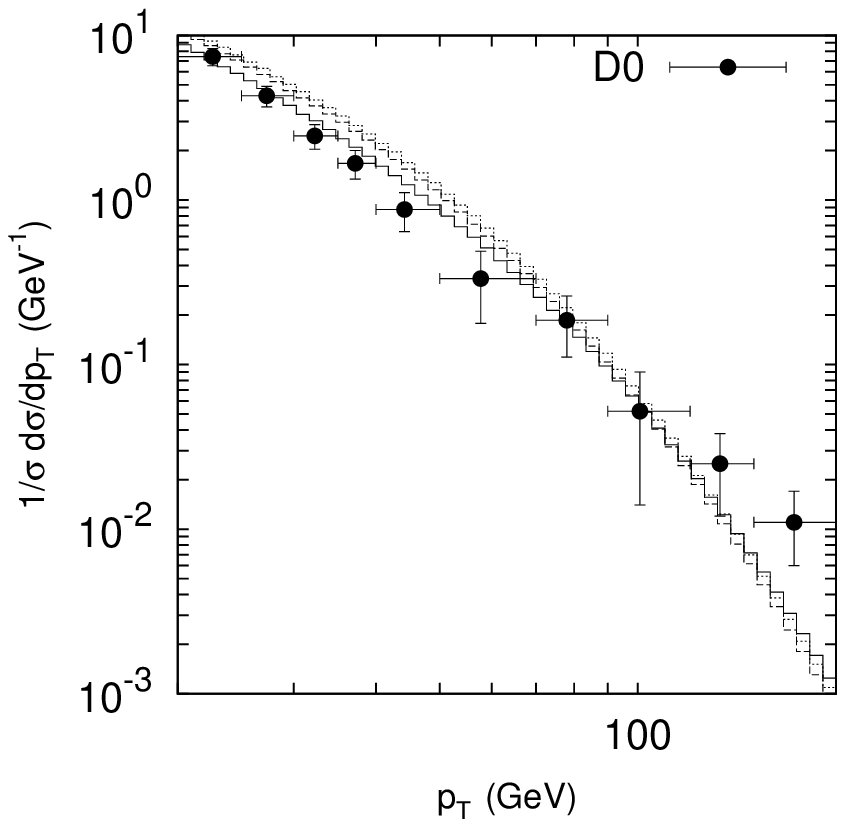, width = 8.1cm}
\caption{Normalized transverse momentum distribution of
the $W^\pm$ boson production calculated at $\sqrt s = 1800$~GeV. Notation of the histograms is 
the same as in Fig.~5. The experimental data are from D$\oslash$~[2].}
\end{center}
\label{figure6}
\end{figure}

\newpage

\begin{figure}
\begin{center}
\epsfig{figure=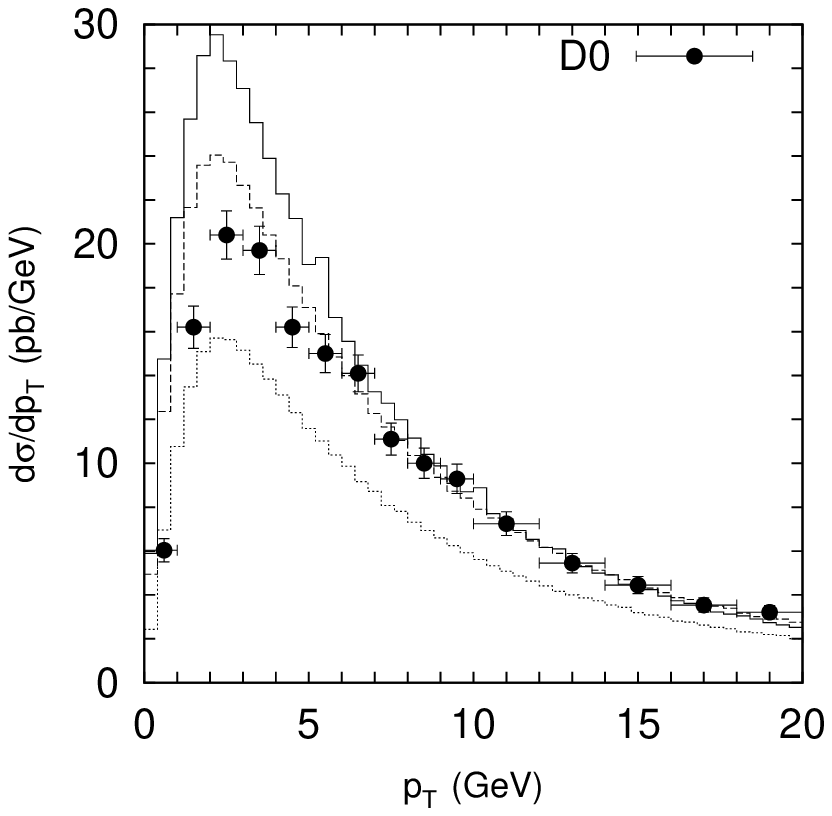, width = 8.1cm}
\epsfig{figure=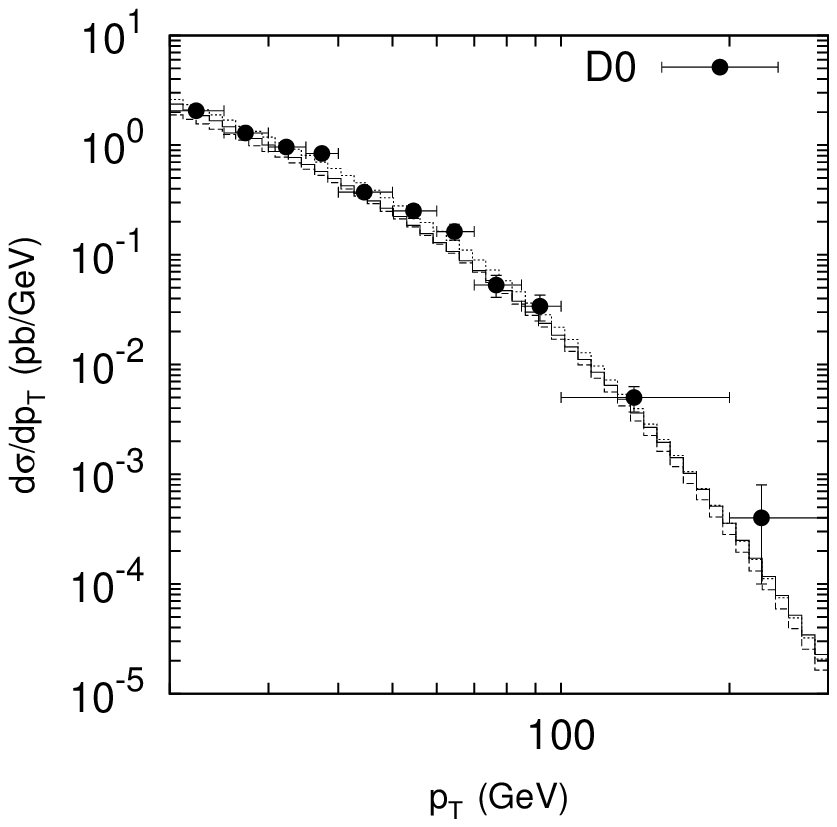, width = 8.1cm}
\epsfig{figure=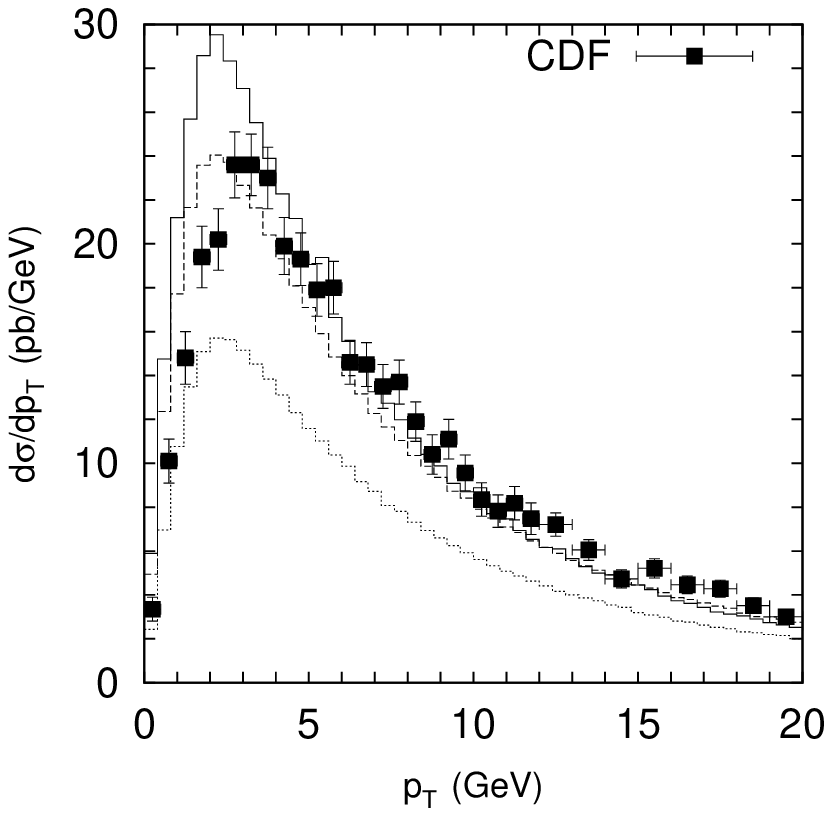, width = 8.1cm}
\epsfig{figure=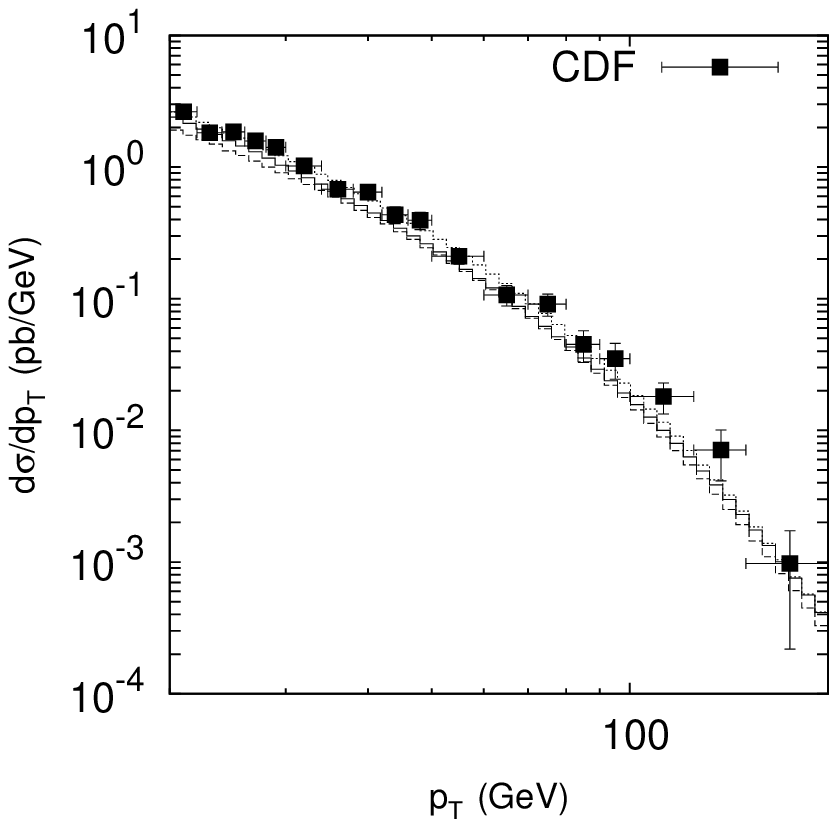, width = 8.1cm}
\caption{Transverse momentum distribution of
the $Z^0$ boson production calculated at $\sqrt s = 1800$~GeV. 
Notation of the histograms is the same as in Fig.~5.
The cross sections time branching fraction $f(Z \to l^+l^-)$ are shown.
The experimental data are from D$\oslash$~[3] and CDF~[1].}
\end{center}
\label{figure7}
\end{figure}

\newpage

\begin{figure}
\begin{center}
\epsfig{figure=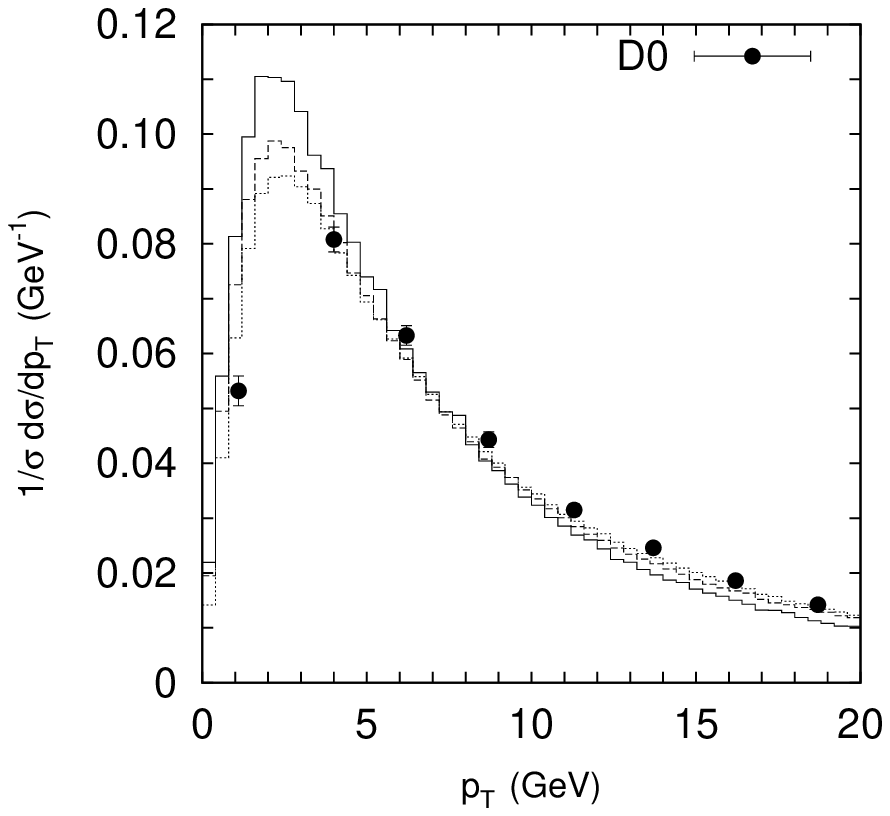, width = 8.1cm}
\epsfig{figure=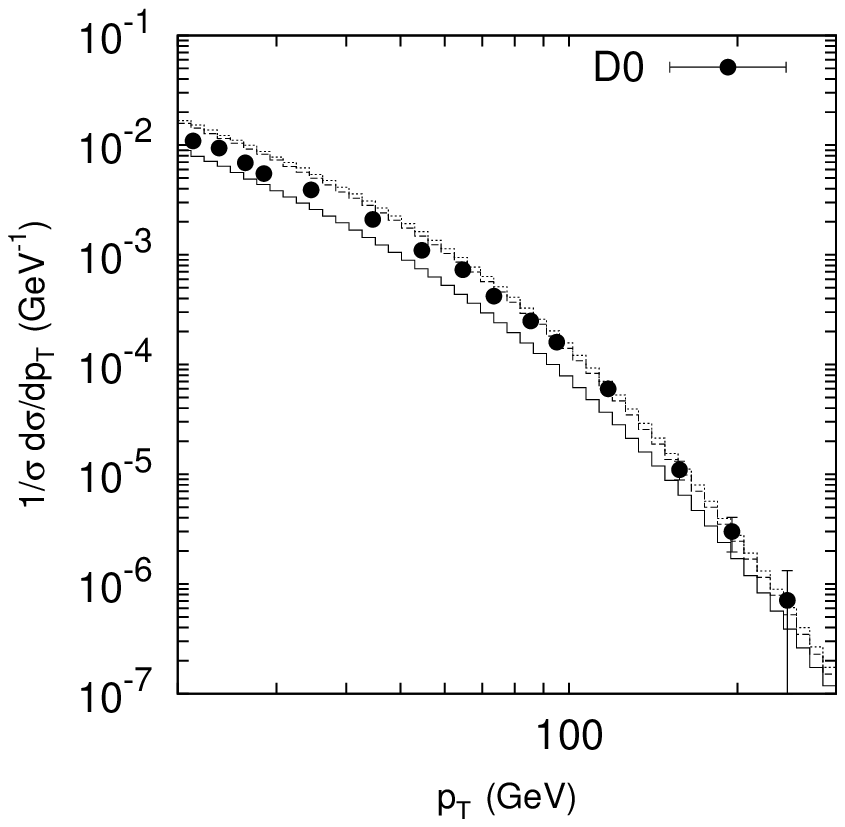, width = 8.1cm}
\caption{Normalized transverse momentum distribution of
the $Z^0$ boson production calculated at $\sqrt s = 1960$~GeV. Notation of the histograms is
the same as in Fig.~5. The experimental data are from D$\oslash$~[6].}
\end{center}
\label{figure8}
\end{figure}

\newpage

\begin{figure}
\begin{center}
\epsfig{figure=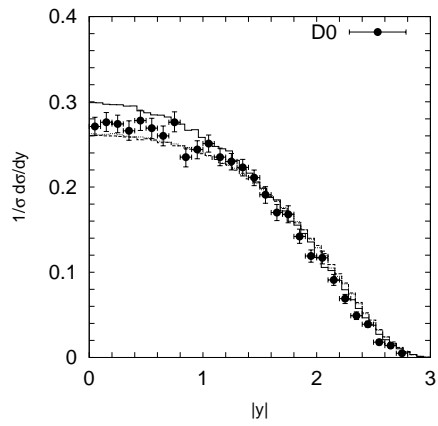, width = 8.1cm}
\caption{Normalized rapidity distribution of
the $Z^0$ boson production calculated at $\sqrt s = 1960$~GeV. Notation of the histograms is 
the same as in Fig.~5. The experimental data are from D$\oslash$~[5].}
\end{center}
\label{figure9}
\end{figure}

\end{document}